\documentstyle[11pt,epsfig]{article}
\title{\bf Corrections to the Cardy-Verlinde formula from the modified dispersion relation in extra dimensions}
\author{A. S. Sefiedgar\thanks{e-mail: a-sefiedgar@sbu.ac.ir} and H. R. Sepangi\thanks{email: hr-sepangi@sbu.ac.ir}
\\ {\small Department of Physics, Shahid Beheshti University, G. C.,  Evin,
Tehran 19839, Iran}}
\begin{document}
\maketitle
\begin{abstract}
The modified dispersion relation as a common feature of all quantum gravity scenarios provides a perturbation framework upon which the black hole thermodynamics can be corrected. In this letter, we obtain the corrections to the $d$-dimensional Schwarzschild black hole thermodynamics by utilizing the extra dimensional form of the modified dispersion relation, leading to the modification of  the Cardy-Verlinde formula. Furthermore, we show that the modified dispersion relation corrections to the Cardy-Verlinde formula can be taken into account by  redefining the Virasoro operator and the central charge.
\end{abstract}
\vspace{2cm}
\section{Introduction}
The Cardy-Verlinde (CV) formula proposed by Verlinde relates the entropy of a certain conformal field theory (CFT) to its total energy
and its Casimir energy in arbitrary dimensions \cite{1}. Using the $\textmd{AdS}_{\textit{d}}/\textmd{CFT}_{\textit{d}-1}$ \cite{2} and $\textmd{dS}_{\textit{d}}/\textmd{CFT}_{\textit{d}-1}$ \cite{3} correspondence, this formula holds exactly for different black holes \cite{4,5,6,7,8,9,10,11,12,13}.
Recently, CV formula was generalized to asymptotically flat spacetimes \cite{14,15} where it was  shown that the $d$-dimensional Schwarzschild solution, which is asymptotically flat, satisfies the generalization of the well known CV formula
\begin{eqnarray}\label{0}
S_{CFT}=\frac{2\pi R}{d-2}\sqrt{2EE_c}
\end{eqnarray}
where the related CFT lives in a $(d-1)$-dimensional spacetime \cite{14,16,17}.

Black holes are suitable examples of extreme quantum gravity regimes which should be described by a complete quantum theory of gravity \cite{18.5}. All promising candidates for quantum gravity expect the existence of a minimal observable length \cite{19,20,21,22,23} and therefore the modified dispersion relation (MDR) is well suitable to explore such a finite resolution of the space-time in the framework of the standard model. MDR is a common feature to all candidates of quantum gravity and, in particular, to the
study of loop quantum gravity (LQG) and of models based on non-commutative geometry. In a similar vein, there has been strong interest in modifications to the energy-momentum dispersion relation \cite{24,25,26,27,28}.
Using MDR to study a black hole thermodynamical behavior and comparing the results with other approaches may further our understanding of their properties and structure.
As other, more fundamental theories such as string theory and loop quantum gravity
may be used to study black hole thermodynamics,  the results of such studies
are useful to impose constraints on the MDR \cite{29} which would ultimately result in a better understanding of quantum gravity. A study along these lines was performed in a previous work \cite{29}.
In addition, the extra dimensional form of the modified dispersion relation introduced in \cite{299}
has also been used in the past to improve our understanding in such studies. In this paper, we
use the extra dimensional form of MDR, obtained in the above mentioned studies where terms proportional to odd powers of energy are not present \cite{29,299,30}, and obtain the corrections to the $d$-dimensional  Schwarzschild black hole thermodynamics by utilizing the MDR. Furthermore,
using such corrected quantities as the entropy, temperature and energy of a black hole, we  derive the modified CV formula within MDR.

Previously, the Schwarzschild black hole had been considered to modify the CV formula from the generalized uncertainty principle \cite{16} and space non-commutativity \cite{18}.
Although the extra dimensional form of the generalized uncertainty principle as an equivalent face of the modified dispersion relation has been used frequently \cite{A1,A2,A3} in the past, the modified dispersion relation in extra dimensions is new.
In this work, we first introduce the modified dispersion relation introduced in \cite{299}. We then use the MDR to obtain the black hole thermodynamics corrections. Knowing the corrections to the black hole thermodynamical quantities, we derive the corrections to the CV formula from MDR. It is  then shown that the corrections to the CV formula may also be derived by just redefining the Virasoro operator and the central charge within MDR.

\section{Extra dimensional form of the modified dispersion relation}
The 4-dimensional form of the modified dispersion relation is \cite{27}
\begin{equation}\label{1}
(\overrightarrow{p})^2=f(E,m;L_p)\simeq E^2-\mu^2+\alpha_1 L_p
E^3+\alpha_2L_p^2E^4+{\cal O}\left(L_P^3E^5\right),
\end{equation}
where $f$ is the function that gives the exact dispersion relation
and $L_p$ is the Planck length. On the right hand we have
used a Taylor-series expansion for $E\ll
\frac{1}{L_p}$. The coefficients $\alpha_i$ may take different values in
other quantum gravity scenarios. Note that $m$ is the rest
energy of the particle and the mass parameter $\mu$ on the right
hand side is directly related to the rest energy. However we note that $\mu\neq m$ if
$\alpha_i$'s do not all vanish. To include quantum gravity effects, the Bekenstein-Hawking theory of black hole thermodynamics has to be modified.  In addition,
loop quantum gravity and string theory give the entropy-area
relation of black holes (for $A\gg L_P^2$)
\begin{equation}\label{2}
S=\frac {A}{4L_P^2}+\rho \ln{\frac{A}{L_p^2}}+{\cal
O}\left(\frac{L_p^2}{A}\right),
\end{equation}
where we expect different values for $\rho$  in string theory and in loop
quantum gravity \cite{27,28,31}. As string theory and loop quantum gravity are believed to provide a more fundamental solution to black
hole thermodynamics, such solutions can be compared to
the ones obtained using the MDR \cite{29}. Therefore, the entropy of
a black hole obtained using equation (\ref{1}) is functionally different from what one obtains using string
theory and loop quantum gravity given by equation (\ref{2}). Introduction of constraints 
on the usual form of the MDR therefore become necessary to obtain a consistent black hole thermodynamics in both approaches.
The result is that terms proportional to odd powers of energy should be ignored in the MDR formula
\cite{29,30}. As a consequence, we take the 4-dimensional form of the MDR as
\begin{equation}\label{3}
(\overrightarrow{p})^2=f(E,m;L_p)\simeq E^2-\mu^2+\alpha
L_p^2E^4+{\cal O}\left(L_P^4E^6\right),
\end{equation}
in what follows. 

Alternatively, using the extra dimensional form of MDR which was investigated in  \cite{299} could be of interest. as is well known,  the generalized uncertainty principle (GUP) is the usual feature of all promising candidates for quantum gravity and implies a minimal observable length in the same way as in MDR. Since the relationship between the generalized uncertainty principle  and MDR is phenomenologically close, one may use the extra dimensional form of GUP, which is known, to find the extra dimensional form of the MDR \cite{299}.  It seems as if the GUP and MDR are phenomenologically two, faces of an underlying quantum gravity proposal. The expectation  that they lead to equivalent results in extra dimensions cannot be considered as far fetched. Now, demanding equivalent forms for GUP and MDR in extra dimensions, one may introduce the $d$-dimensional form of the MDR \cite{299} to modify  thermodynamics of the $d$-dimensional Schwarzschild black hole. Following \cite{299}, we write
\begin{equation}\label{3.5}
(\overrightarrow{p})^2=f(E,m;L_p)\simeq E^2-\mu^2+\alpha
L_p^2E^4+{\cal O}\left(L_P^4E^6\right).
\end{equation}
It is clear that the form of the MDR in extra dimensions is the same as that in 4-dimensions but $L_p$ now depends on the dimensionality of the spacetime to incorporate the effects of the existence of the extra dimensions.

\section{Black hole thermodynamics within MDR}
A $d$-dimensional Schwarzschild black hole is described by the metric
\begin{eqnarray}\label{4}
ds^2=-\left(1-\frac{m}{r^{d-3}}\right)dt^2+\left(1-\frac{m}{r^{d-3}}\right)^{-1}dr^2+r^2d\Omega^2_{d-2}.
\end{eqnarray}
The mass of the black hole is given by $M=\frac{(d-2)\Omega_{d-2}m}{16\pi G_d}$, where $\Omega_{d-2}=\frac{2\pi^{(d-2)/2}}{\Gamma[(d-2)/2]}$ is the area of a unit $(d-2)$ sphere, $d\Omega^2_{d-2}$ is the linear element on the unit sphere $S^{d-2}$ and $G_d$ is Newton's constant in $d$-dimensions.

The locations of outer horizons are written as $r_s=m^{\frac{1}{d-3}}$ \cite{17}.
To find the corrections to black hole thermodynamics within MDR, we start with differentiating equation (\ref{3.5}),
\begin{eqnarray}\label{5}
dp \simeq dE\left[1+\frac{3}{2}\alpha L_p^2E^2+{\cal O}\left(\alpha ^2 L_P^4E^6\right)\right],
\end{eqnarray}
where we have considered only the first order correction terms and the rest mass has been neglected.
We can then write
\begin{eqnarray}\label{6}
dE \simeq dp\left[1-\frac{3}{2}\alpha L_p^2E^2+{\cal O}\left(\alpha ^2 L_P^4E^6\right)\right].
\end{eqnarray}
To first order  corrections, assuming $E\simeq \delta E$, we may apply the standard uncertainty formulae, $\delta E\geq \frac{1}{\delta x}$ and $\delta p\geq\frac{1}{\delta x}$, to obtain \cite{29,299,30}
\begin{eqnarray}\label{7}
dE\geq \frac{1}{\delta x}\left[1-\frac{3}{2}\alpha L_p^2\frac{1}{{\delta x}^2}+{\cal O}\left(\alpha ^2 L_P^4\frac{1}{{\delta x}^6}\right)\right].
\end{eqnarray}
The corrected energy of the black hole within MDR may be written as $E'$ which is assumed to be $E'\simeq dE$
\begin{eqnarray}\label{7.5}
E'\geq \frac{1}{\delta x}\left[1-\frac{3}{2}\alpha L_p^2\frac{1}{{\delta x}^2}+{\cal O}\left(\alpha ^2 L_P^4\frac{1}{{\delta x}^6}\right)\right].
\end{eqnarray}
In the process of Hawking radiation, the uncertainty in the position of
a Hawking particle at the emission is $\delta x\simeq 2r_s$ \cite{29,299}.
Utilizing  equation $\delta x\simeq 2m^{\frac{1}{d-3}}$, we  have
\begin{eqnarray}\label{8}
E' \simeq \frac{1}{2m^{\frac{1}{d-3}}}-\frac{3}{16}\alpha L_p^2\frac{1}{m^{\frac{3}{d-3}}}+{\cal O}\left(\alpha ^2 L_P^4\frac{1}{m^{\frac{5}{d-3}}}\right).
\end{eqnarray}
Now, we only consider the first order correction term without loss of generality and write the black hole modified energy as $E'=E+\Delta E$, where
\begin{eqnarray}\label{9}
E \simeq \frac{1}{2m^{\frac{1}{d-3}}}    \quad    ,   \quad     \Delta E \simeq -\frac{3}{16}\alpha L_p^2\frac{1}{m^{\frac{3}{d-3}}}.
\end{eqnarray}
Note that $E$ is the standard energy and the presence of MDR in a quantum gravity regime results in the appearance of $\Delta E$.
Now, since the Hawking temperature can be identified with  energy \cite{16}, we set the constant of proportionality to $\frac{d-3}{2\pi}$ and get
\begin{eqnarray}\label{10}
T' \simeq \frac{d-3}{4\pi}\frac{1}{m^{\frac{1}{d-3}}}-\frac{3(d-3)}{32\pi} \alpha L_p^2 \frac{1}{m^{\frac{3}{d-3}}}+{\cal O}\left(\alpha ^2 L_P^4\frac{1}{m^{\frac{5}{d-3}}}\right).
\end{eqnarray}
Considering only the first order correction term, one may write the temperature as $T'=T+\Delta T$, where
\begin{eqnarray}\label{11}
T \simeq \frac{d-3}{4\pi}\frac{1}{m^{\frac{1}{d-3}}}   \quad , \quad   \Delta T \simeq -\frac{3(d-3)}{32\pi} \alpha L_p^2 \frac{1}{m^{\frac{3}{d-3}}}.
\end{eqnarray}
Note that $T$ is the standard black hole temperature and the emergence of $\Delta T$ is due to the effects of MDR in quantum gravity regime.
Using the first law of thermodynamics, one may write
\begin{eqnarray}\label{11.5}
dS'_{d} \simeq \frac{\Omega_{d-2}}{4G_d}{\frac{(d-2)}{(d-3)}} dm \left[m^{\frac{1}{d-3}}+\frac{3}{8}\alpha L_p^2 m^{\frac{-1}{d-3}}+{\cal O}\left(\alpha ^2 L_P^4 m^{\frac{-3}{d-3}}\right)\right].
\end{eqnarray}
Thus, the entropy of the black hole can be written as

\begin{eqnarray}\label{12}
S'_{d\neq 4} \simeq \frac{\Omega_{d-2}}{4G_d}m^{\frac{d-2}{d-3}}+\frac{3\Omega_{d-2}(d-2)}{32G_d(d-4)}\alpha L_p^2 m^{\frac{d-4}{d-3}},
\end{eqnarray}
which is true for all $d$'s except $d=4$ up to the first order  correction term.
It is important to point out that if one tries to consider higher order correction terms, the entropy for odd $d$ and even $d$ will be different. In fact, a logarithmic correction term will appear in the entropy formula of a black hole with even dimensions for higher order correction terms and this is due to the presence of the factors $m^{\frac{-3}{d-3},\frac{-5}{d-3},\frac{-7}{d-3}...}$ in equation (\ref{11.5}).

Let us now write the entropy as $S'=S+\Delta S$, where
\begin{eqnarray}\label{13}
S \simeq  \frac{\Omega_{d-2}}{4G_d}m^{\frac{d-2}{d-3}}   \quad , \quad   \Delta S_{d\neq 4} \simeq \frac{3\Omega_{d-2}(d-2)}{32G_d(d-4)}\alpha L_p^2 m^{\frac{d-4}{d-3}}.
\end{eqnarray}
For $d=4$, the entropy of a black hole can be written as
\begin{eqnarray}\label{14}
S'_{d=4} \simeq S+ \frac{3\Omega_2}{16 G_4} \alpha L_p^2 \ln m,
\end{eqnarray}
where
\begin{eqnarray}\label{15}
\Delta S_{d=4} \simeq  \frac{3\Omega_2}{16 G_4} \alpha L_p^2 \ln m.
\end{eqnarray}
We again note that $S$ is the standard black hole entropy and the emergence of $\Delta S$ is due to the effects of MDR in a quantum gravity regime.
Therefore, the black hole thermodynamic corrections are obtained from the MDR. It is important to stress that we have considered only the first order correction term during the calculations, without any loss of generality.

As an important point, we mention the appearance of a logarithmic correction term in the entropy of a 4-dimensional black hole.
In the recent past, many researchers have become interested in fixing the coefficient of the logarithmic correction term within the statistical and quantum geometrical approaches \cite{B4,B5,B6,B7,B8,B9,B10,B11}. Since, this particular parameter
might be useful as a discriminator of prospective fundamental theories, fixing it independent of the specific elements of any one particular model of quantum gravity seems to be of importance. Therefore, MDR as a model independent concept is a suitable approach to provide the corrections to  black hole entropy.

As has been argued in  \cite{29}, the parameter $\alpha$ in MDR is a negative quantity of order one (see also \cite{30}). There, a comparison was made between the results of two approaches, the generalized uncertainty principle and modified dispersion
relation within the context of black hole thermodynamics with that of string theory and Loop
quantum gravity. Demanding the same results in all approaches and considering string theory and
loop quantum gravity as more comprehensive, some constraints were imposed on the form of GUP and
MDR. Also, it was found that GUP and MDR are not independent concepts. In fact, they could be
equivalent in an ultimate quantum theory of gravity. The existence of a positive minimal observable
length necessitates a positive value for the model dependent parameter $\alpha$ in the form of GUP. Since we know the relation between the model dependent parameters in GUP and MDR in \cite{29}, we set the
parameter $\alpha$ as a negative value for MDR in this paper.
Now it is easy to see that the logarithmic correction term in 4-dimensional black hole entropy is negative in our approach.
In spite of the current lack of detailed knowledge about $\alpha$, our calculation might still be
considered as a step towards a final theory of quantum gravity. The fundamental
theory may be able to make a precise statement about $\alpha$ and, as a result, a prediction about the coefficient of the logarithmic term in the present calculation. It may also have something to say about the logarithmic
coefficient through more direct means. The correspondence of the results via these two sets of calculations can be a good test for the validity of the final quantum gravity theory \cite{B6}.

Furthermore, it is interesting to note that the existence of
a logarithmic correction term in the entropy relation is restricted to even black hole dimensionality within MDR.
In fact if one tries to consider the higher order correction terms, the emergence of a logarithmic term in the entropy relation for even dimensions will be certain.  Of course, the presence of a logarithmic correction term will put a constraint on the dimensionality of the black hole and its dual CFT.  It would also result in a better insight when dealing with formulating quantum gravity.

\section{MDR corrections to the CV formula}
The entropy of a (1+1)-dimensional CFT is given by the well known Cardy formula
\begin{eqnarray}\label{16.1}
S_{CFT}=2\pi \sqrt{\frac{c}{6}\left(L_0-\frac{c}{24}\right)},
\end{eqnarray}
where $L_0=ER$ is the product of energy and radius and the shift  $\frac{c}{24}$ is caused by the Casimir effect \cite{32}. After making the appropriate identifications for $L_0$ and $c$, the same Cardy formula is also valid for CFT in arbitrary $(d-1)$-dimensional spacetime in the form \cite{1}
\begin{eqnarray}\label{16.2}
S_{CFT}=\frac{2\pi R}{d-2} \sqrt{E_c(2E-E_c)},
\end{eqnarray}
where $R$ is the radius of the system, $E$ is the total energy and $E_C$ is the Casimir energy defined as
\begin{eqnarray}\label{16.3}
E_c=(d-1)E-(d-2)TS.
\end{eqnarray}
Recently, the CV formula was generalized to asymptotically flat spacetimes (the generalized CV formula) \cite{14,15}. In an asymptotically flat spacetime, it is given by \cite{14,16,17}
\begin{eqnarray}\label{17}
S_{CFT}=\frac{2\pi R}{d-2}\sqrt{2E E_c}.
\end{eqnarray}
We have computed the modified dispersion relation corrections to the thermodynamics of a $d$-dimensional Schwarzschild black hole described by the generalized CV formula (\ref{17}). Therefore, we can write the CV formula within MDR for arbitrary $d$ as
\begin{eqnarray}\label{17.2}
S'_{CFT}=\frac{2\pi R}{d-2}\sqrt{2E' E'_c}
\end{eqnarray}
and the Casimir energy, equation (\ref{16.3}), within MDR as
\begin{eqnarray}\label{18}
E'_c=(d-1)E'-(d-2)T'S',
\end{eqnarray}
where a prime indicates the quantity in MDR. Using equation (\ref{18}), we substitute $E'=E+\Delta E$, $T'=T+\Delta T$ and $S'=S+\Delta S$ in equation (\ref{17.2}) to obtain
\begin{eqnarray}\label{19}
S'_{CFT} \simeq S_{CFT}\left[1+\frac{(d-2)}{2EE_c}\left(\frac{2(d-1)}{(d-2)}E\Delta E-ET\Delta S-E\Delta T S-\Delta E T S\right)\right].
\end{eqnarray}
Since the correction terms induced by MDR are small, we have ignored their product and used Taylor expansion. Substituting the quantities in terms of $E$, one may obtain the modified dispersion relation corrections of CV formula as
\begin{eqnarray}\label{20}
S'_{CFT} \simeq S_{CFT}\left\{1+\frac{(d-2)\alpha L_p^2}{2E E_c}\left[  -\frac{3(d-1)}{(d-2)}E^4 +  \frac{3}{256 \pi} \frac{(d-3) (d-6)}{(d-4)}\frac{\Omega_{d-2}}{G_d} (2E)^{6-d}  \right]\right\},
\end{eqnarray}
which is valid for $d \neq 4$ up to  first order correction terms.
It is also important to write the corrections to CFT entropy for $d=4$ as
\begin{eqnarray}\label{21}
S'_{CFT} \simeq S_{CFT}\left[1+\frac{\alpha L_p^2 }{E E_c}\left(  -\frac{9}{2}E^4 +  \frac{3}{32 \pi}\frac{\Omega_2}{G_4}E^2 (\ln{2E}+1)  \right)\right].
\end{eqnarray}
The correction terms are certainly caused by the modified dispersion relation.
For  Schwarzschild black holes, the dual CFT lives in a flat space and the energy has no subextensive part \cite{14,16}. Since the Casimir energy vanishes \cite{14}, the CV formula (\ref{16.2}) makes no sense in this case \cite{14,16}.

In 2-dimensional conformal field theory, when the conformal weight of the ground state is zero, the CV formula
\begin{eqnarray}\label{22}
S=2\pi \sqrt{\frac{cL_0}{6}},
\end{eqnarray}
is valid \cite{14,16,33}. Note that $c$ is the central charge and $L_0$ is the Virasoro operator.
If we use $E_cR=(d-2)\frac{S_c}{2\pi}$ in equation (\ref{16.2}), where $S_c$ is the Casimir entropy and drop  $E_c$ in analogy with equation (\ref{22}), we obtain the generalization to equation (\ref{22}) in $(d-1)$ dimensional CFT \cite{14,16,33} as
\begin{eqnarray}\label{22.5}
S=\frac{2\pi}{d-2} \sqrt{\frac{cL_0}{6}},
\end{eqnarray}
 where
\begin{eqnarray}\label{23}
L_0=RE   \quad , \quad   \frac{c}{6}=\frac{(d-2)S_c}{\pi}=2E_cR.
\end{eqnarray}
Now equation (\ref{22.5}) can be applied to a Schwarzschild black hole with asymptotically flat spacetime.
In fact, as was noted before,  the CV (the generalized CV) formula is the outcome of a striking resemblance between the thermodynamics of CFTs with asymptotically Ads (flat) duals and CFTs in two dimensions \cite{14,16,33}.
Therefore, it is possible to take into account the MDR corrections to  CV entropy formula by just redefining the Virasoro operator and the central charge, the quantities entering the CV formula. The Virasoro operator can be modified to
\begin{eqnarray}\label{24}
L'_0=RE'=L_0-\frac{3}{2}\alpha L_p^2 R E^3,
\end{eqnarray}
which is correct for all values of $d$. In this manner, the central charge can be written as
\begin{eqnarray}\label{25}
c'_{d \neq 4}=c+12\alpha L_p^2 R \left[-\frac{3 (d-1)}{2}E^3-\frac{3}{64\pi}\frac{(d-3)(d-2)}{(d-4)}\frac{\Omega_{d-2}}{G_d} (2E)^{5-d} \right],
\end{eqnarray}
and
\begin{eqnarray}\label{26}
c'_{d=4}=c+12\alpha L_p^2 R \left[\frac{-9}{2}E^3+\frac{3}{16\pi}\frac{\Omega_2}{G_4} E \left(\ln(2E)+\frac{1}{2}\right) \right].
\end{eqnarray}
Therefore, one may consider the standard form of the CV formula, but redefine the Virasoro operator and central charge to take into account the MDR corrections.

\section{Conclusions}
We have computed the modified dispersion relation corrections to the entropy of a $d$-dimensional Schwarzschild black hole described by the CV formula, equation (\ref{17}). We have used the modified dispersion relation to correct the thermodynamical properties of the $d$-dimensional Schwarzschild black hole such as entropy, temperature and energy. The corrected black hole thermodynamical quantities have been used to modify the CV formula. In other words, we have obtained the corrections to the entropy of a dual conformal field theory living in a flat space.  We also stressed the point that the CV (generalized CV) formula is the outcome of an interesting resemblance between the thermodynamics of CFTs with asymptotically Ads (flat) duals and CFTs in two dimensions. We then derived the corrections to the Virasoro operator and the central charge, the quantities entering the CV formula. We have shown the possibility of taking into account the MDR corrections to the CV entropy formula by just redefining the Virasoro operator and the central charge.

\end{document}